\documentclass[aps,preprint]{revtex4}
\usepackage{color}
\usepackage{graphicx}
\usepackage{amsmath}
\usepackage{amssymb}
\usepackage[]{hyperref}
\usepackage{bm}
\usepackage{ulem}

\usepackage{color} 

\begin{document}

\title{Quantum fidelity of electromagnetically induced transparency: The full quantum theory}

\author{Hao Hsu$^{1}$, Chin-Yao Cheng$^{1}$, Jiun-Shiuan Shiu$^{1}$, Ling-Chun Chen$^{1}$, and Yong-Fan Chen$^{1,2}$}

\email{yfchen@mail.ncku.edu.tw}

\affiliation{$^1$Department of Physics, National Cheng Kung University, Tainan 70101, Taiwan \\
$^2$Center for Quantum Technology, Hsinchu 30013, Taiwan
}



\begin{abstract}
We present a full quantum model to study the fidelity of single photons with different quantum states propagating in a medium exhibiting electromagnetically induced transparency (EIT). By using the general reservoir theory, we can calculate the quantum state of the transmitted probe photons that reveal the EIT phenomenon predicted by semiclassical theory while reflecting the influence of the quantum fluctuations of the strong coupling field. Our study shows that the coupling field fluctuations not only change the quantum state of the probe photons, but also slightly affect its transmittance. Moreover, we demonstrate that the squeezed coupling field can enhance the influence of its fluctuations on the quantum state of the probe photons, which means that the EIT effect can be manipulated by controlling the quantum state properties of the coupling field. The full quantum theory in this paper is suitable for studying quantum systems related to the EIT mechanism that would allow us to examine various quantum effects in EIT-based systems from a full quantum perspective.
\end{abstract}


\pacs{42.50.Gy, 42.65.Ky, 03.67.-a, 32.80.Qk}


\maketitle


\newcommand{\FigOne}{
    \begin{figure}[t]
    \centering
    \includegraphics[width=11.0cm]{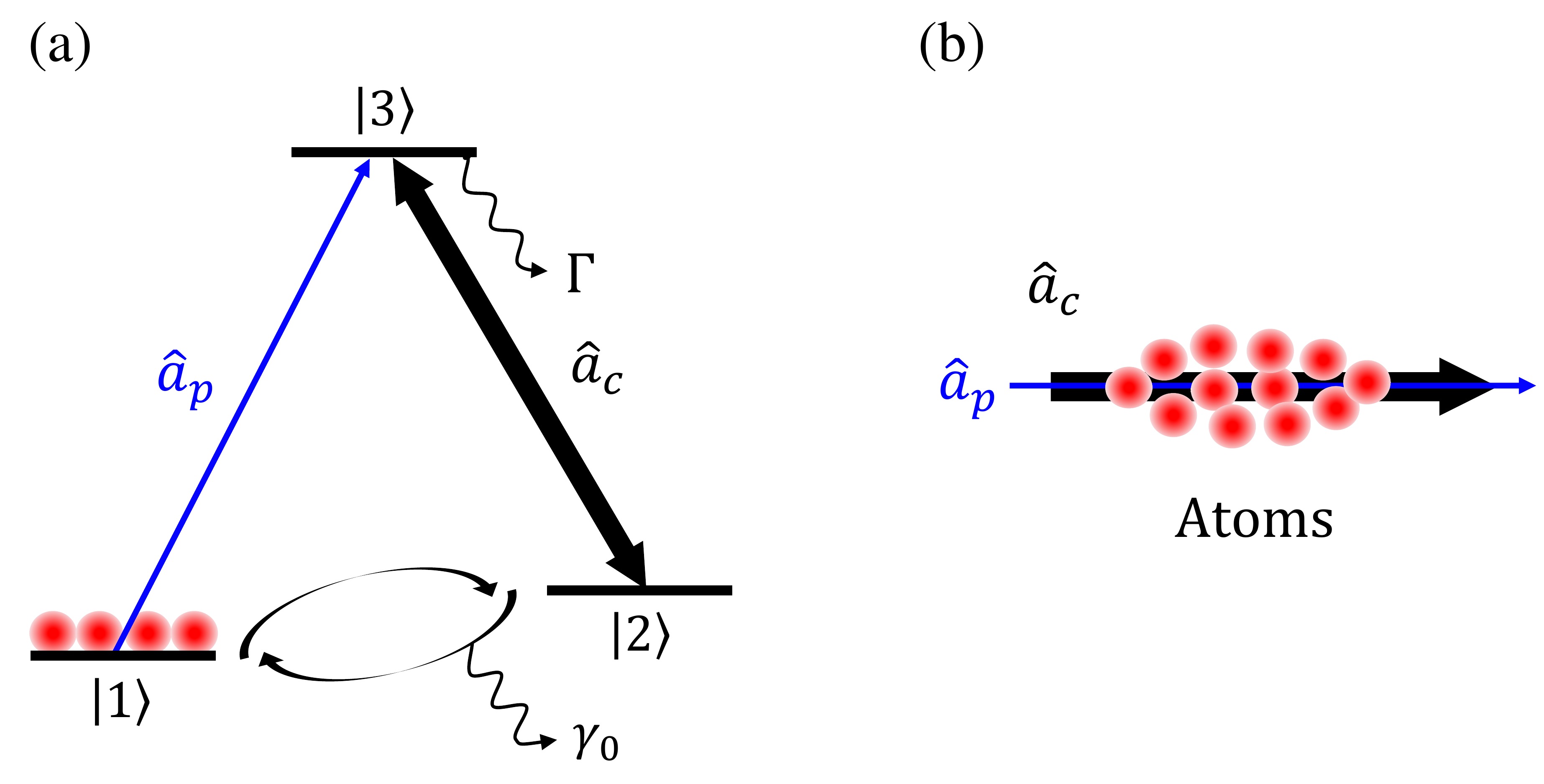}
    \caption{
(a) Schematic of the $\Lambda$-type EIT system, where $\hat{a}_p$ and $\hat{a}_c$ represent the annihilation operators of the weak probe field and strong coupling field, respectively. (b) The two quantum fields interact with the atomic ensemble.
}
    \end{figure}
}


\newcommand{\FigTwo}{
    \begin{figure}[t]
    \centering
    \includegraphics[width=7.5cm]{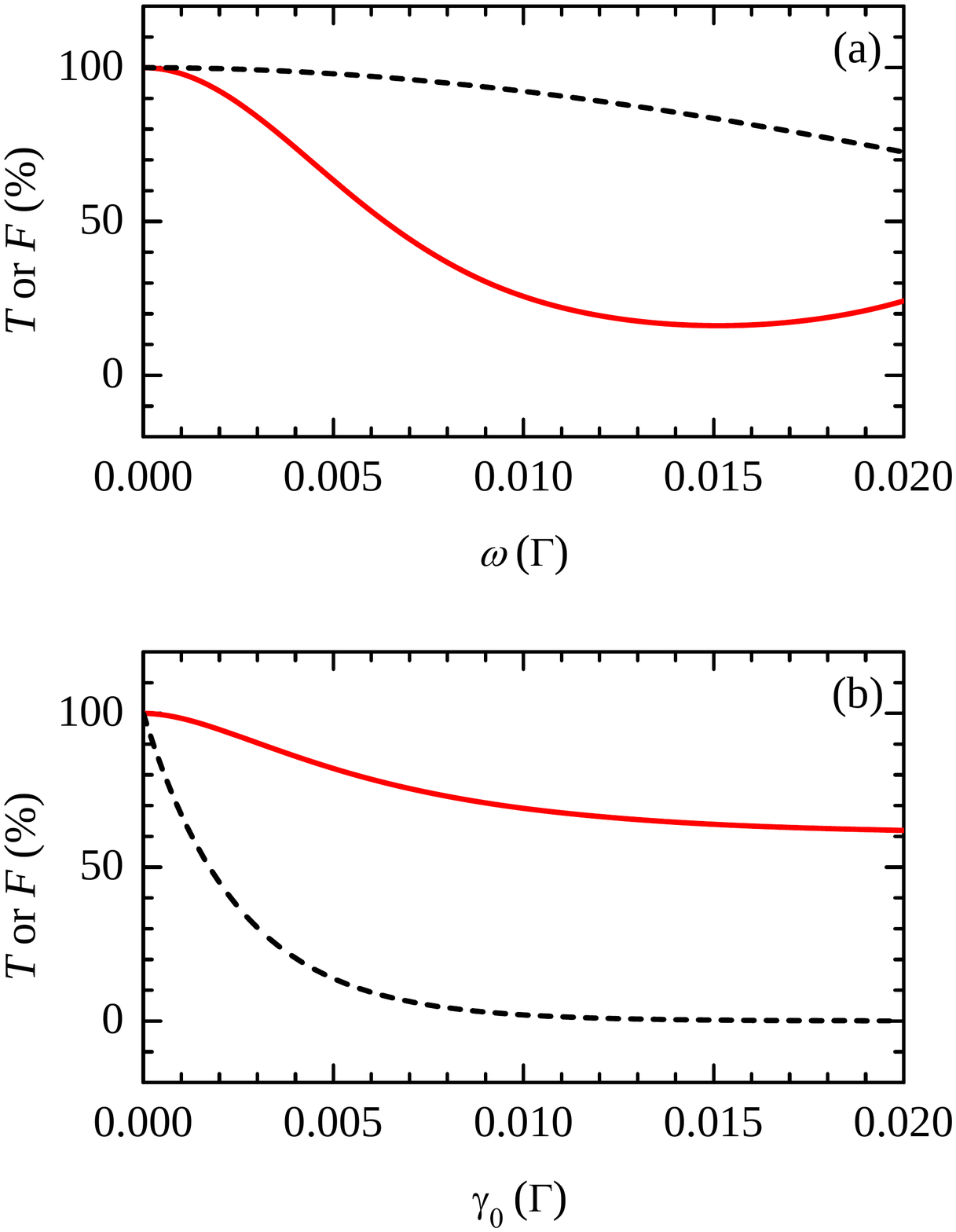}
    \caption{
The transmittance $T$ (black dashed lines) and fidelity $F$ (red solid lines) of the single-photon coherent state vary with the (a) two-photon detuning $\omega$ and (b) ground-state dephasing $\gamma_0$ under the EIT conditions of $\Omega_c = 0.50\Gamma$ and $\alpha = 200$.
}
    \end{figure}
}


\newcommand{\FigThree}{
    \begin{figure}[t]
    \centering
    \includegraphics[width=7.5cm]{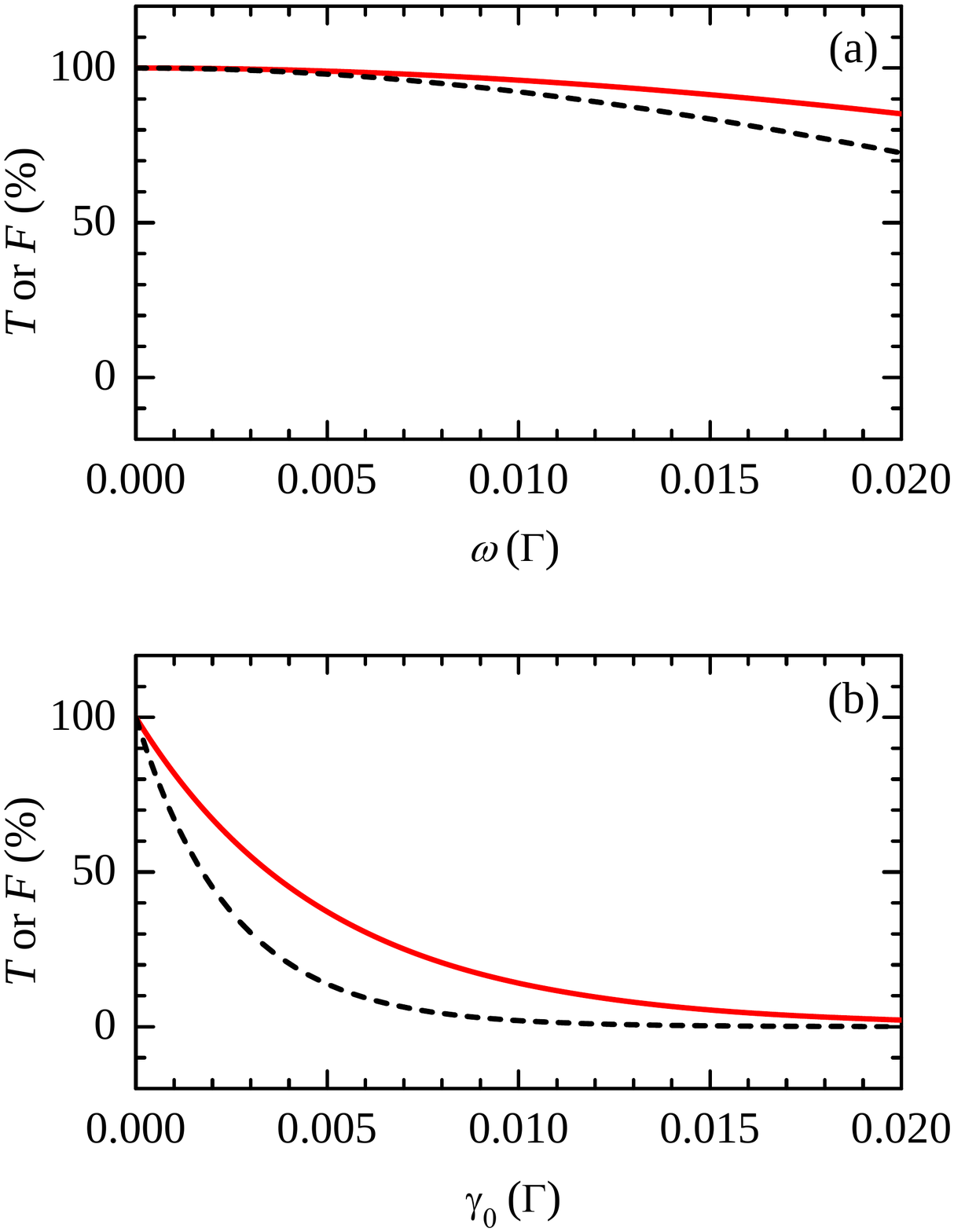}
    \caption{
The transmittance $T$ (black dashed lines) and fidelity $F$ (red solid lines) of the single-photon Fock state vary with the (a) two-photon detuning $\omega$ and (b) ground-state dephasing $\gamma_0$ under the EIT conditions of $\Omega_c = 0.50\Gamma$ and $\alpha = 200$.
}
    \end{figure}
}


\newcommand{\FigFour}{
    \begin{figure}[t]
    \centering
    \includegraphics[width=7.5cm]{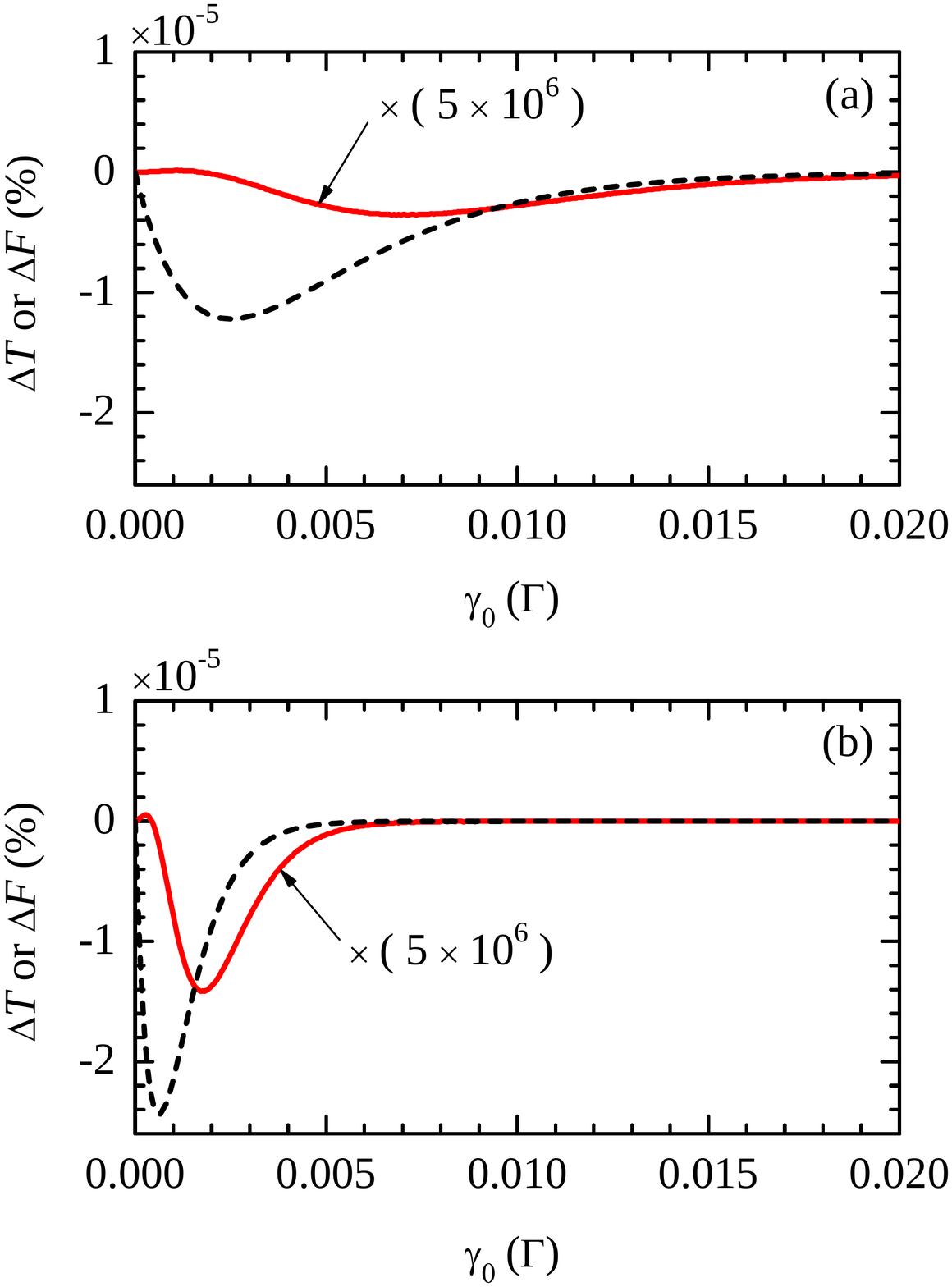}
    \caption{
The transmittance difference $\Delta T$ (black dashed lines) and fidelity difference $\Delta F$ (red solid lines) of the single-photon coherent state vary with the ground-state dephasing $\gamma_0$ under the EIT conditions of (a) $\Omega_c = 0.50\Gamma$ and (b) $\Omega_c = 0.25\Gamma$, respectively. The optical depth $\alpha$ of the medium is set to 200.
}
    \end{figure}
}


\newcommand{\FigFive}{
    \begin{figure}[t]
    \centering
    \includegraphics[width=7.5cm]{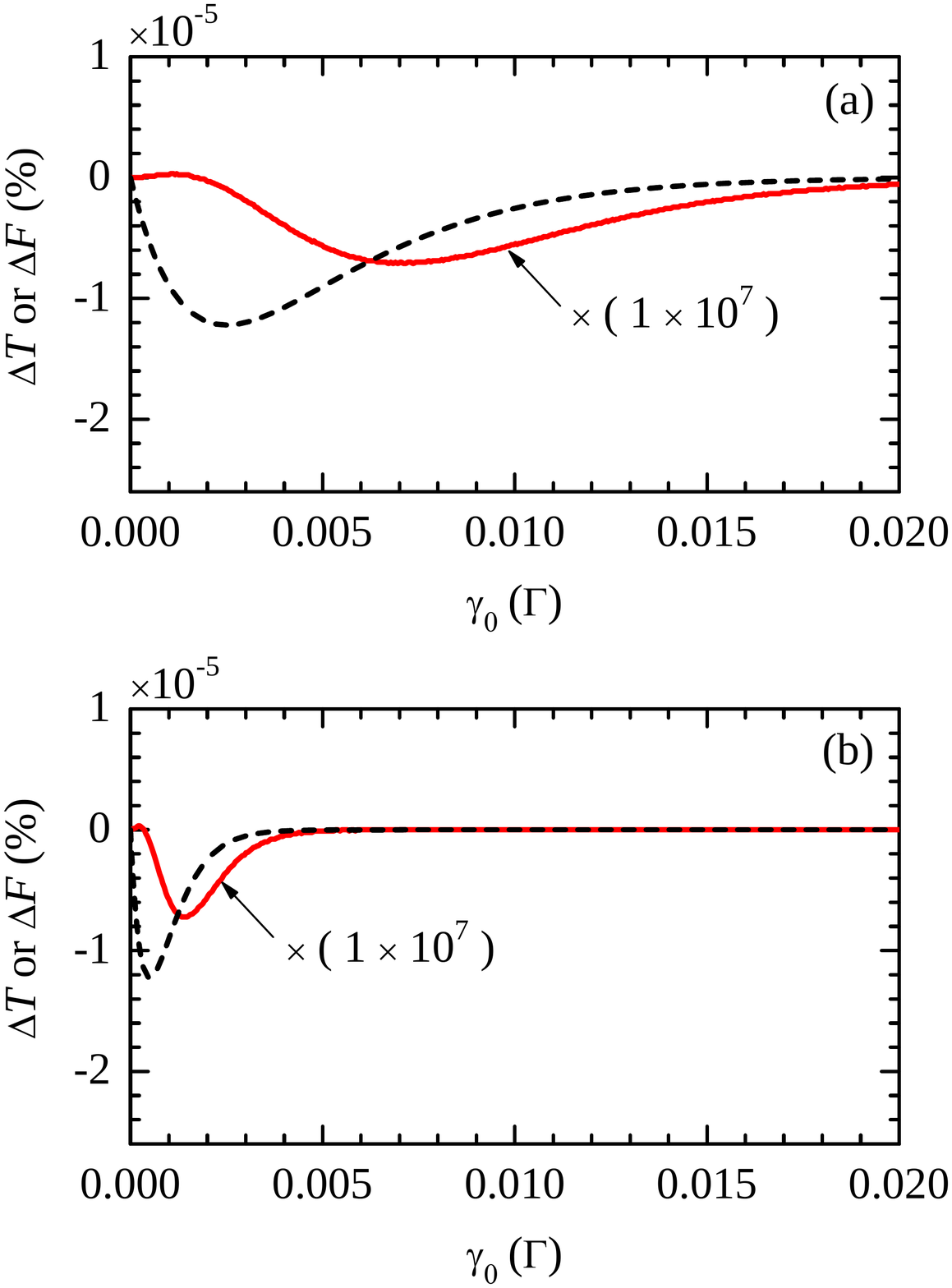}
    \caption{
The transmittance difference $\Delta T$ (black dashed lines) and fidelity difference $\Delta F$ (red solid lines) of the single-photon coherent state vary with the ground-state dephasing $\gamma_0$ under the EIT conditions of (a) $\alpha=200$ and (b) $\alpha=1000$, respectively. Here, the coupling Rabi frequency $\Omega_c$ is set to $0.50\Gamma$.
}
    \end{figure}
}


\newcommand{\FigSix}{
    \begin{figure}[t]
    \centering
    \includegraphics[width=7.5cm]{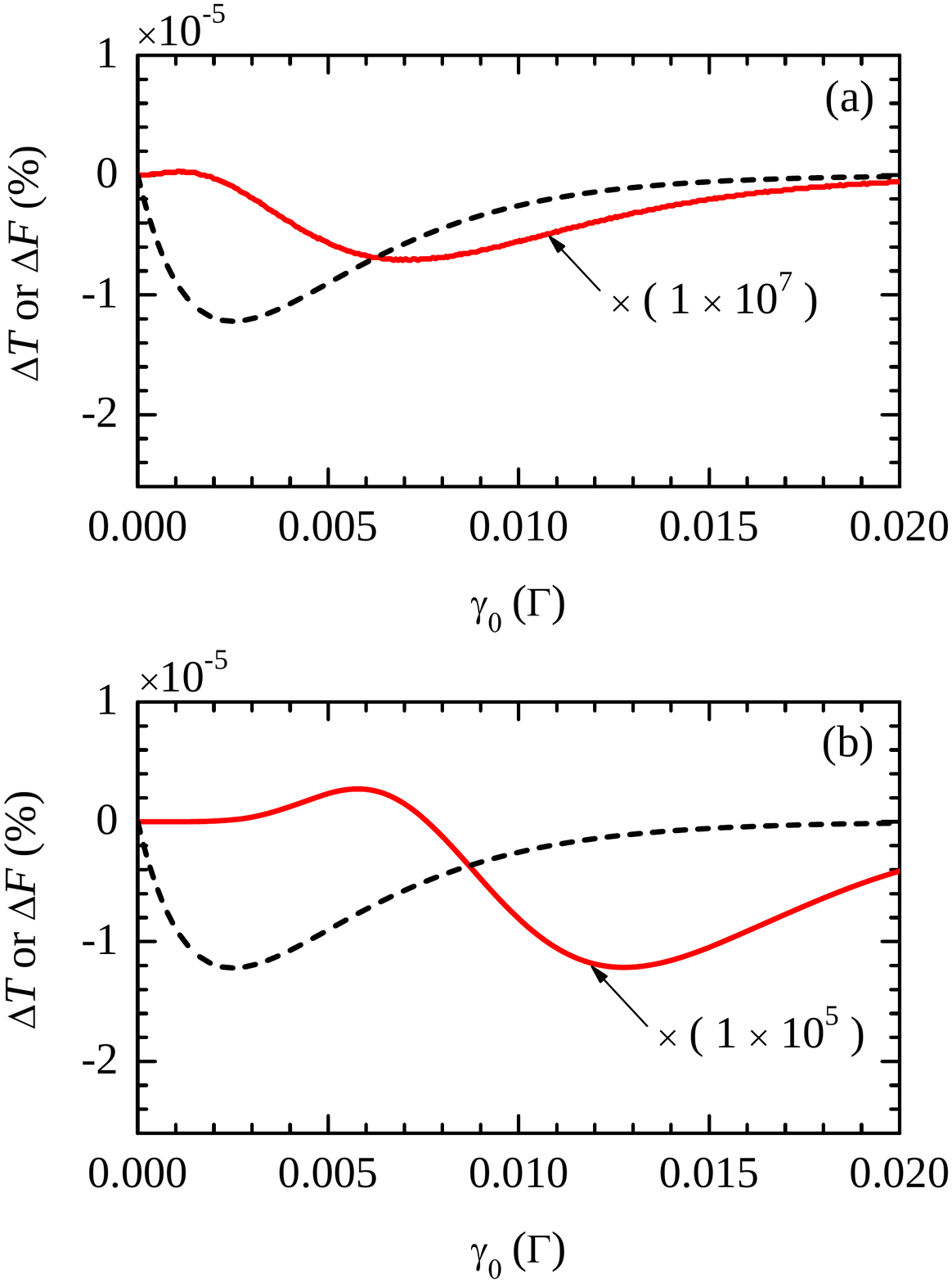}
    \caption{
The transmittance difference $\Delta T$ (black dashed lines) and fidelity difference $\Delta F$ (red solid lines) under different probe photon number conditions. The mean photon number of the coherent probe photons are (a) $n_{p0}=1$ and (b) $n_{p0}=10$, respectively. The optical depth $\alpha$ and the coupling Rabi frequency $\Omega_c$ are set to 200 and $0.50\Gamma$, respectively.
}
    \end{figure}
}

\section{Introduction}

Both photons and atoms can carry quantum information and are promising for the development of quantum computing and quantum communication. Manipulating the various properties of carried quantum states (e.g., entanglement and squeezing) and developing well-characterized qubits and quantum logic components through atom--photon interactions are essential topics in quantum information science~\cite{LukinDSP, FleischhauerDSP, LukinQG, LukinDB, DLCZ}. In particular, the coherent and reversible storage of the quantum states of photons is a core technology for executing numerous protocols in quantum information processing~\cite{KimbleQI, PolzikQI, GisinQR, BlattQI, RempeQM}. In the past two decades, substantial research has pursued such photon storage by leveraging the phenomenon of electromagnetically induced transparency (EIT)~\cite{HarrisEIT, FleischhauerEIT, HauSL,ScullySL,HauLS, LukinSL, SqueezedEIT, LukinQM, YuXPM, LukinLS, LvovskyQM, YuLS, ChenLS, ZhuQM, LauratQM}. Because the EIT medium with high dispersion can significantly enhance the nonlinear interaction at the single-photon level and suppress the vacuum field noise, many EIT-based quantum applications such as photonic transistors~\cite{PTransister1, PTransister2, PTransister3}, optical phase gates~\cite{XPM1, XPM2, XPM3, ChenXPM}, and quantum frequency converters~\cite{YuFBS, ChenQFC1, ChenQFC2} have been proposed and demonstrated. The latest developments in EIT systems based on waveguide, circuit, and cavity electrodynamics are also worth noting~\cite{QED EIT1, QED EIT2, QED EIT3, QED EIT4}.

In a common EIT system, the strong coupling field makes the otherwise opaque medium transparent near the atomic resonance; thus, the weak probe field can propagate without absorption. Early studies suggested that the quantum state of the probe photons combines with the quantum state of the atoms to form a dark-state polariton in EIT~\cite{FleischhauerDSP}. This dark-state polariton is not affected by the vacuum field reservoir under ideal conditions, and can convert the quantum state between photons and atoms by changing the strength of the coupling field~\cite{FleischhauerQM, DantanQM, XiaoDSP}. Therefore, EIT can be used for effective optical quantum memory, which can also protect photons' entangled states, even if ground-state dephasing is considered~\cite{PengQM, KimbleQM}. However, according to the literature review, the current theory of the EIT dark-state polariton not only lacks the research on the fidelity of photons with different quantum states propagating in the EIT medium, but also does not consider the quantum properties of the coupling field~\cite{FleischhauerQFWM}. This investigation centers on establishing a full quantum model to calculate the EIT fidelity of the probe photons with different quantum states, and study the influence of the quantized coupling field on the EIT medium.

We present a full quantum model based on mean-field expansion~\cite{FeischhauerQEIT, DaviddovischQF, YangEntanglement} and the general reservoir theory~\cite{ScullyQO} to study the quantum properties of the EIT medium. Using the proposed full quantum model, we confirm that under the influence of ground-state dephasing or two-photon detuning, the fidelity of Fock-state photons and coherent-state photons propagating in the EIT medium are quite different. Moreover, we demonstrate that the quantum fluctuations of the coupling field not only change the quantum state of the probe photons, but also slightly affect its transmittance. The remainder of this paper is organized as follows. In Section~\ref{sec:TM}, we describe the full quantum model of EIT and theoretically determine the probe photons' transmittance under the influence of the vacuum field reservoir and coupling field fluctuations. In Section~\ref{sec:Fidelity}, we apply the general reservoir theory to calculate the evolution of the wave function of probe photons propagating in the EIT medium. The quantum fidelity of the single-photon coherent state and Fock state under the influence of EIT are both obtained. Finally, in Section~\ref{sec:CF}, we calculate the contribution of the coupling field fluctuations to the fidelity and transmittance of the probe photons. In the study of the influence of coupling field fluctuations on the EIT fidelity, the coupling field strength, number of probe photons, and optical depth of the EIT medium are the variables considered.

\FigOne


\section{Theoretical Model}
\label{sec:TM}

We study a typical $\Lambda$-type EIT system from a full quantum perspective, as shown in Fig. 1. In the three-level atomic system, the ground states $|1\rangle$ and $|2\rangle$ are coupled to the excited state $|3\rangle$ by using a weak probe field and a strong coupling field, respectively. Here, we consider that the propagation directions of the two light fields are the same to eliminate the influence of the Doppler frequency shift in the atomic medium. Both the weak probe field ($\hat{a}_p$, $|1\rangle\leftrightarrow|3\rangle$ transition) and strong coupling field ($\hat{a}_c$, $|2\rangle\leftrightarrow|3\rangle$) are described in a quantized manner as
\begin{eqnarray}
\hat{E}_{p(c)}=\sqrt{\frac{\hbar \omega_{p(c)}}{2\epsilon_0 V}} \hat{a}_{p(c)}e^{-i\omega_{p(c)}t + ik_{p(c)} z}+\textrm{c.c.},
\end{eqnarray}
where $k_{p(c)}$, $\omega_{p(c)}$, and $\hat{a}_{p(c)}$ represent the wave vector, angular frequency, and annihilation operator of the probe (coupling) field, respectively; $V$ is the interaction volume; and $\epsilon_0$ is the permittivity in a vacuum. The parameter $\gamma_0$ in Fig. 1 represents the dephasing rate of the two ground states $|1\rangle$ and $|2\rangle$. $\Gamma$ is the spontaneous decay rate of the excited state $|3\rangle$ contributed by the vacuum field reservoir. To simplify the theoretical model, we assume that the decay rate from the excited state $|3\rangle$ to the ground state $|1\rangle$ is the same as that from the same excited state to the ground state $|2\rangle$. Therefore, if $\gamma_{j}$ represents the rate of decay from the excited state $|3\rangle$ to the ground state $|j\rangle$, then $\gamma_1 = \gamma_2 = 0.5 \Gamma \equiv \gamma$.

When the probe field is considerably weaker than the coupling field, the probe field can be regarded as a perturbation field in a medium exhibiting EIT. To study the propagation behavior of probe photons in the medium, we use the Heisenberg--Langevin equations and Maxwell--Schr\"{o}dinger equation to solve the first-order perturbation equations as follows:
\begin{align}
&\frac{\partial}{\partial t}\hat{\sigma}^{}_{13}=-\gamma\hat{\sigma}^{}_{13}+ig_{p}\hat{a}_p+ig_{c}\hat{a}_c\hat{\sigma}^{}_{12}+\hat{F}_{13},\\
&\frac{\partial}{\partial t}\hat{\sigma}^{}_{12}=-\gamma_0 \hat{\sigma}^{}_{12}+ig_{c}\hat{a}^{\dagger}_c\hat{\sigma}^{}_{13}+\hat{F}_{12},\\
&\bigg(\frac{\partial}{\partial t}+c\frac{\partial}{\partial z}\bigg)\hat{a}_p=ig_{p}^{*}N\hat{\sigma}^{}_{13},
\end{align}
where $\hat{\sigma}_{\mu\nu}$ is a collective atomic operator in the slowly varying amplitude approximation and $\hat{F}_{\mu\nu}$ denotes the associated Langevin noise operator~\cite{FleischhauerQM}. $N$ is the total number of atoms in the medium; $c$ is the speed of light in a vacuum; and $g_{p(c)}=\frac{d_{31(32)}\varepsilon_{p(c)}}{\hbar}$ is the coupling constant between the probe (coupling) field and the atom, where $d_{\mu\nu}$ is the dipole moment of the corresponding transition and $\varepsilon_{p(c)}=\sqrt{\frac{\hbar \omega_{p(c)}}{2\epsilon_0 V}}$ is the electric field of the single probe (coupling) photon.

To solve Eqs. (2)--(4), we apply a method of mean-field expansion similar to that used in \cite{FeischhauerQEIT, DaviddovischQF, YangEntanglement}, expressing the operators as steady-state mean values plus fluctuation operators. We first solve for the steady-state mean values and revise Eqs. (2)--(4) accordingly:
\begin{align}
&0=-\gamma\langle\hat{\sigma}_{13}\rangle+ig\langle\hat{a}_p\rangle+i\Omega_c\langle\hat{\sigma}_{12}\rangle,\\
&0=-\gamma_0 \langle\hat{\sigma}_{12}\rangle+i\Omega^{*}_c\langle\hat{\sigma}_{13}\rangle,\\
&c\frac{\partial}{\partial z}\langle \hat{a}_p\rangle=ig^{*}N\langle\hat{\sigma}_{13}\rangle,
\end{align}
where we assume that $g_p = g_c = g,$ and where the semiclassical form of the Rabi frequency of the coupled field $\Omega_c= g\langle\hat{a}_c\rangle$ is used to simplify the theoretical model. Under a weak probe field approximation, the amount of energy lost in the propagation of the strong coupling field in the EIT medium is negligible~\cite{FleischhauerQM}. Thus, $\Omega_c$ can be regarded as a constant in the propagation along the \textit{z-}axis. If the medium length is $L$, the solutions of Eqs. (5)--(7) at $z=L$ are
\begin{align}
&\langle\hat{a}_{pL}\rangle={\rm{exp}}\bigg(-\frac{|g|^{2}N}{c}\frac{\gamma_0}{\gamma_0 \gamma+|\Omega_{c}|^2}L\bigg)\langle\hat{a}_{p0}\rangle,\\
&\langle\hat{\sigma}_{13}\rangle=\frac{ig\gamma_0}{\gamma_0 \gamma+|\Omega_{c}|^2}\langle\hat{a}_{pL}\rangle,\\
&\langle\hat{\sigma}_{12}\rangle=\frac{i\Omega^{*}_c}{\gamma_0}\langle\hat{\sigma}_{13}\rangle,
\end{align}
where $\hat{a}_{pL}=\hat{a}_p(z=L)$ and $\hat{a}_{p0}=\hat{a}_p(z=0)$ for simplification. Substituting Eq. (9) into Eq. (10), the steady-state solution of $\langle\hat{\sigma}_{12}\rangle$ under the condition of strong coupling field is approximately $-g\langle\hat{a}_{pL}\rangle / \Omega_{c}$. Next, we apply mean-field expansion to express the operators:
\begin{align}
&\hat{a}_{c}=\langle\hat{a}_{c}\rangle+\delta \hat{a}_{c},\\
&\hat{\sigma}_{\mu\nu}=\langle\hat{\sigma}_{\mu\nu}\rangle+\delta\hat{\sigma}_{\mu\nu}.
\end{align}
By incorporating Eqs. (11) and (12) into Eqs. (2)--(4) and using the steady-state results obtained from Eqs. (5)--(10),  the equations of motion can be rewritten as follows:
\begin{align}
&\frac{\partial}{\partial t}\delta\hat{\sigma}_{13}=-\gamma\delta\hat{\sigma}_{13}-ig\langle\hat{a}_p\rangle+ig\hat{a}_p+ig\langle\hat{\sigma}_{12}\rangle\delta\hat{a}_c+i\Omega_c\delta\hat{\sigma}_{12}+\hat{F}_{13},\\
&\frac{\partial}{\partial t}\delta\hat{\sigma}_{12}=-\gamma_0 \delta\hat{\sigma}_{12}+i\Omega^{*}_c\delta\hat{\sigma}_{13}+ig\langle\hat{\sigma}_{13}\rangle\delta\hat{ a}^{\dagger}_c+\hat{F}_{12},\\
&\bigg(\frac{\partial}{\partial t}+c\frac{\partial}{\partial z}\bigg)\hat{a}_p=ig^{*}N(\langle\hat{\sigma}_{13}\rangle+\delta\hat{\sigma}_{13}).
\end{align}
Note that due to the strong coupling field conditions $\langle\hat{a}_{c}\rangle \gg \delta\hat{a}_c$, the higher-order fluctuation terms in Eqs. (13) and (14) have been ignored~\cite{ChuangQE}. Subsequently, by applying the Fourier transform $\tilde Q (z, \omega)=\frac{1}{\sqrt{2\pi}}\int\hat{Q}(z,t)e^{i\omega t}dt$ to Eqs. (13)--(15), the spatial differential equation for the probe field is obtained:
\begin{align}
&\frac{\partial}{\partial z}\tilde{a}_{p}=-\kappa_{a}\tilde{a}_{p}-e^{-\Lambda_{0}z}\kappa_{b}\delta\tilde{ a}_{c}-e^{-\Lambda_{0}z}\kappa_{c}\delta\tilde{ a}^{\dagger}_{c}+\tilde n.
\end{align}
This equation indicates that the spatial evolution of the propagation of the probe field in the medium is affected by the quantum fluctuations of the coupling field. The relevant parameters are listed as follows:
\begin{align}
&\Lambda_{0}=\frac{|g|^{2}N\gamma_{0}}{c(\gamma_{0}\gamma+|\Omega_{c}|^{2})},\\
&\kappa_{a}=\frac{|g|^{2}N(\gamma_{0}-i\omega)}{c[(\gamma_0-i\omega)(\gamma-i\omega)+|\Omega_{c}|^{2}]}-\frac{i\omega}{c},\\
&\kappa_{b}=\frac{|g|^{2}N(\gamma_{0}-i\omega)}{c[(\gamma_0-i\omega)(\gamma-i\omega)+|\Omega_{c}|^{2}]}\frac{\Omega^{*}_{c}}{(\gamma_{0}\gamma+|\Omega_{c}|^{2})}g\langle \tilde{a}_{p0}\rangle,\\
&\kappa_{c}=\frac{|g|^{2}N\gamma_0}{c[(\gamma_0-i\omega)(\gamma-i\omega)+|\Omega_{c}|^{2}]}\frac{\Omega_{c}}{(\gamma_{0}\gamma+|\Omega_{c}|^{2})}g\langle \tilde{a}_{p0}\rangle,\\
&\tilde{n}=-\frac{g^{*}N}{c}\frac{\Omega_c}{(\gamma_0-i\omega)(\gamma-i\omega)+|\Omega_c|^{2}}\tilde{F}_{12}+\frac{ig^{*}N}{c}\frac{\gamma_0-i\omega}{(\gamma_0-i\omega)(\gamma-i\omega)+|\Omega_c|^{2}}\tilde{F}_{13},
\end{align}
where the parameter $\tilde{n}$ contains all the Langevin noise terms. The parameter $\kappa_a$ describes the dispersion characteristics of the EIT medium, which appears only in the first term on the right-hand side of Eq. (16) and can be derived using the semiclassical model~\cite{PengQM}. A nonzero dephasing rate $\gamma_0$ or probe detuning $\omega$ gives $\kappa_a \neq 0$, indicating the attenuation of the propagation of the probe photons through the EIT medium. Here, the frequency of the coupling field is set to resonance with the medium; thus, $\omega$ is also indicative of the two-photon detuning of the EIT medium. The parameters $\Lambda_0$, $\kappa_b$, and $\kappa_c$ represent the influences of coupling field fluctuations on the probe photons, as presented in Eq. (16). The appearance of $\Lambda_0$ in the exponential function indicates the exponential decay of the quantum fluctuation effect in the EIT medium. In addition, $\kappa_b$ and $\kappa_c$ represent the influence coefficients of the coupling field fluctuations $\delta\tilde{a}_c$ and $\delta\tilde{a}_c^\dagger$, respectively, and may attenuate the propagation of the probe photons in the EIT medium. These results also indicate that the influence of the coupling field fluctuations on the probe photons must be accompanied by its dissipation in the EIT medium, which can come from nonzero ground-state dephasing and two-photon detuning. However, under the condition $\Omega_c\gg g \langle\tilde{a}_{p0}\rangle$, both $\kappa_b$ and $\kappa_c$ are close to zero, which means that an extremely strong coupling field can suppress the influence of the quantum fluctuations on the probe photons.

We solve Eq. (16) under the initial condition of $\tilde a_{p}(L=0, \omega)=\tilde{a}_{p0}(\omega)$. The general solution for the probe photons is
\begin{align}
\tilde{a}_{pL}(\omega)=e^{-\kappa_{a}L}\tilde{a}_{p0}(\omega)+\frac{e^{-\kappa_{a}L}-e^{-\Lambda_{0}L}}{\kappa_{a}-\Lambda_0}(\kappa_{b}\delta\tilde{ a}_{c}+\kappa_{c}\delta\tilde{ a}^{\dagger}_{c})+\frac{1-e^{-\kappa_{a}L}}{\kappa_{a}}\tilde{n}(L, \omega).
\end{align}
This solution is indeterminate at $\omega=0$, where $\kappa_a$ is equal to $\Lambda_0$, making the second term on the right-hand side 0/0. For the special case of $\omega=0$, we set $\Lambda_0=\kappa_a$ as an additional initial condition of Eq. (16). The solution becomes
\begin{align}
\tilde{a}_{pL}(0)=e^{-\kappa_{a}L}\tilde{a}_{p0}(0)-e^{-\kappa_{a}L}L(\kappa_{b}\delta\tilde{ a}_{c}+\kappa_{c}\delta\tilde{ a}^{\dagger}_{c})+\frac{1-e^{-\kappa_{a}L}}{\kappa_a}\tilde{n}(L, 0).
\end{align}
For simplicity, the general solution of the probe photons propagating in the EIT medium can be expressed as
\begin{align}
\tilde{a}_{pL}(\omega)=c_1\tilde{a}_{p0}(\omega)+\tilde{c_2}+\tilde{f},
\end{align}
where the following are the relevant parameters:
\begin{equation}
 c_1 = e^{-\kappa_{a}L},
\end{equation}
\begin{equation}
\tilde{c_2} =
        \begin{cases}
        \frac{e^{-\kappa_{a}L}-e^{-\Lambda_{0}L}}{\kappa_{a}-\Lambda_0}(\kappa_{b}\delta\tilde{ a}_{c}+\kappa_{c}\delta\tilde{ a}^{\dagger}_{c}) & \text{if $\omega\neq 0$},\\
        -e^{-\kappa_{a}L}L(\kappa_{b}\delta\tilde{ a}_{c}+\kappa_{c}\delta\tilde{ a}^{\dagger}_{c}) & \text{if $\omega=0$},
        \end{cases}
\end{equation}
\begin{equation}
\tilde{f} = \frac{1-e^{-\kappa_{a}L}}{\kappa_a}\tilde{n}(L, \omega).
\end{equation}
Given that the mean number of probe photons propagating in the EIT medium can be obtained from $n_p(z, \omega) = \langle\tilde{a}_p^\dagger(z, \omega)\tilde{a}_p(z, \omega)\rangle$, by using Eq. (24), the mean photon number in the output probe field is obtained:
\begin{align}
\begin{split}
n_{pL}&=\langle\tilde{a}_{pL}^\dagger\tilde{a}_{pL}\rangle\\
&=\langle(c_1^*\tilde{a}_{p0}^\dagger+\tilde{c}_2^\dagger+\tilde{f}^\dagger)(c_1\tilde{a}_{p0}+\tilde{c}_2+\tilde{f})\rangle,
\end{split}
\end{align}
where $\tilde{f}$ contains two Langevin noise operators $\tilde{F}_{12}$ and $\tilde{F}_{13}$. This explains the vacuum reservoir--related appearance of the terms 
$\langle\tilde{a}_{p0}^\dagger\tilde{F}_{\mu\nu}\rangle$,
$\langle\tilde{F}_{\mu\nu}^\dagger\tilde{a}_{p0}\rangle$,
$\langle\delta\tilde{a}_c^\dagger\tilde{F}_{\mu\nu}\rangle$,
$\langle\tilde{F}_{\mu\nu}^\dagger\delta\tilde{a}_c\rangle$, and
$\langle\tilde{F}_{\mu\nu}^\dagger\tilde{F}_{\mu'\nu'}\rangle$  in $n_{pL}$ , where $\mu\nu\in\{12, 13\}$ and $\mu'\nu'\in\{12, 13\}$. Because the input probe field and the coupling field fluctuations are both statistically independent of the vacuum reservoir, the terms $\langle\tilde{a}_{p0}^\dagger\tilde{F}_{\mu\nu}\rangle$,
$\langle\tilde{F}_{\mu\nu}^\dagger\tilde{a}_{p0}\rangle$,
$\langle\delta\tilde{a}_c^\dagger\tilde{F}_{\mu\nu}\rangle$, and
$\langle\tilde{F}_{\mu\nu}^\dagger\delta\tilde{a}_c\rangle$ are zero. Moreover, the noise correlations $\langle\tilde{F}_{\mu\nu}^\dagger\tilde{F}_{\mu'\nu'}\rangle$ can be expressed as
\begin{align}
\langle\tilde{F}_{\mu\nu}(z, \omega)\tilde{F}_{\nu'\mu'}(z', \omega')\rangle=\frac{L}{2\pi N}D_{\mu\nu,\nu'\mu'}\delta(\omega-\omega')\delta(z-z'),
\end{align}
where $D_{\mu\nu,\nu'\mu'}$ is the diffusion coefficient of the EIT system and can be obtained using the Einstein relation~\cite{ChenQFC1, GarrisonQO}. The delta function $\delta(\omega-\omega')$ describes the short-memory approximation of  the vacuum reservoir modes, and $\delta(z-z')$ appears because we assume that each atom is coupled to only its own reservoir. Notably, because the diffusion coefficient is close to zero under the weak field perturbation condition of the EIT system, the contribution of the Langevin noise term $\tilde{f}$ in Eq. (28) is negligible~\cite{ChenQFC1}. Therefore, the transmittance of the probe photons propagating in the EIT medium is given by
\begin{align}
T=\frac{n_{pL}}{n_{p0}}=\frac{\langle(c_1^*\tilde{a}_{p0}^\dagger+\tilde{c}_2^\dagger)(c_1\tilde{a}_{p0}+\tilde{c}_2)\rangle}{\langle\tilde{a}_{p0}^\dagger\tilde{a}_{p0}\rangle}.
\end{align}
As mentioned earlier, in the case of an extremely strong coupling field ($\Omega_c\gg g \langle\tilde{a}_{p0}\rangle$), both $\kappa_b$ and $\kappa_c$ are close to 0, which means that the influence of the coupling field fluctuations can be ignored ($\tilde{c_2}\approx0$). Thus, the probe photons' transmittance becomes $|c_1|^2 = e^{-2\kappa_{a}L}$, consistent with semiclassical model~\cite{FleischhauerQM, DantanQM, XiaoDSP, PengQM}.


\section{EIT Quantum Fidelity}
\label{sec:Fidelity}

To quantify the influence of the coupling field fluctuations on the propagation of single photons in an EIT medium, we use the general reservoir theory to calculate the fidelity of the wave function of the propagating single photons with different quantum states. We first write the output density matrix of the EIT system as follows:
\begin{align}
\rho_f = \tilde{U} \rho_i \tilde{U}^{\dagger},
\end{align}
where $\rho_i=\rho^{P}(z=0,\omega) \otimes \rho^{C}\otimes \rho^{R}$ represents the initial density matrix containing the probe field, coupling field, and vacuum reservoir in the frequency domain. The unitary matrix $\tilde{U}$ is the evolution operator of the combined system. The density matrix element of the output probe field in the Fock-state basis is given by
\begin{align}
\begin{split}
\rho^{P}_{mn}(L, \omega)&=\langle m|\rho^{P}(L, \omega)|n\rangle\\
&=\langle m|{\rm{Tr}}_{C, R}(\tilde{U}\rho_{i}\tilde{U}^{\dagger})|n\rangle\\
&={\rm{Tr}}_{P}\{|n\rangle\langle m|{\rm{Tr}}_{C, R}(\tilde{U}\rho_{i}\tilde{U}^{\dagger})\}\\
&={\rm{Tr}}\{|n\rangle\langle m|\otimes I_{C}\otimes I_{R} \tilde{U}\rho_{i}\tilde{U}^{\dagger}\}\\
&={\rm{Tr}}\{(\tilde{U}^{\dagger}|n\rangle\langle m|\otimes I_{C}\otimes I_{R}\tilde{U})\rho_{i}\}\\
&={\rm{Tr}}\{\tilde{\rho}_{mn}(L, \omega)\rho_{i}\},
\end{split}
\end{align}
where ${\rm{Tr}}_P$, ${\rm{Tr}}_C$, and ${\rm{Tr}}_R$ are the traces over the degrees of freedom of the probe photons,  coupling field, and vacuum reservoir, respectively. ${\rm{Tr}}$ represents the total trace of the combined system. $I_C$ and $I_R$ are the identity operators in the Hilbert space of the coupling field and vacuum reservoir, respectively. Thus, the operator of the density matrix element $\rho^{P}_{mn}(L,\omega)$ in Eq. (32) is given by
\begin{align}
\tilde{\rho}_{mn}(L,\omega) = \tilde{U}^{\dagger} (\left| n\right\rangle \left\langle m\right|\otimes I_{C} \otimes I_{R}) \tilde{U}.
\end{align}
By treating the Fock state as the photon creation operator acting on the vacuum state $|n\rangle=\frac{(\tilde{a}_{p0}^\dagger )^{n}}{\surd{n!}}|0\rangle$ and by rewriting the vacuum density matrix with the annihilation and creation operators $|0\rangle\langle 0|= \sum_{l=0}^{\infty} \frac{(-1)^l}{l!}(\tilde{a}_{p0}^\dagger)^l(\tilde{a}_{p0})^l$~\cite{LouisellQO}, the operator $\tilde{\rho}_{mn}(L,\omega)$ can be rewritten as
\begin{align}
\begin{split}
\tilde{\rho}_{mn}(L, \omega)&=\frac{1}{\sqrt{m!n!}}\sum_{l=0}^{\infty}\frac{(-1)^l}{l!}(\tilde{a}_{pL}^\dagger)^{l+n}(\tilde{a}_{pL})^{l+m}\\
&=\sum_{l=0}^{\infty}\chi_{mnl}(\tilde{a}_{pL}^\dagger)^{l+n}(\tilde{a}_{pL})^{l+m},
\end{split}
\end{align}
where $\chi_{mnl}=\frac{1}{\sqrt{m!n!}}\frac{(-1)^l}{l!}$. Next, by using Eq. (24) and ignoring the Langevin noise term $\tilde{f}$, the output density matrix element of the probe photons is given by
\begin{align}
\begin{split}
\rho_{mn}^{P}(L, \omega)&={\rm{Tr}}\{\tilde{\rho}_{mn}(L,\omega)\rho_i\}\\
&=\sum_{l=0}^{\infty}\chi_{mnl}{\rm{Tr}}\{(c_1^{*}\tilde{a}_{p0}^{\dagger}+\tilde{c}_2^{\dagger})^{l+n}(c_1\tilde{a}_{p0}+\tilde{c}_2)^{l+m}\rho_i\}.
\end{split}
\end{align}

We now consider two input probe photon states: the coherent state and Fock state. For the case of the coherent state, the density matrix of the input probe photons is expressed as $\rho^P(0,\omega)=|\beta_p\rangle\langle\beta_p|$. In addition, the coupling field in the coherent state $\rho^C=|\beta_c\rangle\langle\beta_c|$ is applied. According to Eq. (35), the density matrix element of the output probe photons propagating in the EIT medium becomes
\begin{align}
\begin{split}
\rho_{mn}^{P}&(L, \omega)\\
&={\rm{Tr}}\{\tilde{\rho}_{mn}(L,\omega)\rho_i\}\\
&=\sum_{l=0}^{\infty}\chi_{mnl}{\rm{Tr}}_{P, C}\{(c_1^{*}\tilde{a}_{p0}^{\dagger}+\tilde{c}_2^{\dagger})^{l+n}(c_1\tilde{a}_{p0}+\tilde{c}_2)^{l+m}|\beta\rangle\langle\beta|\}\\
&=\sum_{l=0}^{\infty}\chi_{mnl}\langle\beta|(c_1^{*}\tilde{a}_{p0}^{\dagger}+\tilde{c}_2^{\dagger})^{l+n}(c_1\tilde{a}_{p0}+\tilde{c}_2)^{l+m}|\beta\rangle\\
&=\sum_{l=0}^{\infty}\chi_{mnl}\langle\beta|(c_1^{*}\beta_p^*+\tilde{c}_2^{\dagger})^{l+n}(c_1\beta_p+\tilde{c}_2)^{l+m}|\beta\rangle,
\end{split}
\end{align}
where $|\beta\rangle=|\beta_p\rangle\otimes|\beta_c\rangle$ for simplification. If the operator $\tilde{c}_2$ related to the coupling field fluctuations in Eq. (36) is ignored, the state of the output probe field, $|c_1\beta_p\rangle$, remains coherent; otherwise, it becomes a mixed state. This indicates that the quantum fluctuations of the coherent coupling field destroy the coherence of the coherent probe photons. To obtain the fidelity of the coherent probe photons propagating through the EIT medium, we employ a processing method similar to that in Eqs. (32)--(34) as follows:
\begin{align}
\begin{split}
F=&\sqrt{|\langle\beta_p|\rho^P(L,\omega)|\beta_p\rangle|}\\
=&\sqrt{|{\rm{Tr}}\{(\tilde{U}^{\dagger}|\beta_p\rangle\langle\beta_p|\otimes I_{C}\otimes I_{R}\tilde{U})\rho_i\}|}\\
=&\sqrt{|{\rm{Tr}}\{\tilde{U}^{\dagger}\tilde{D}(\beta_p)|0\rangle\langle0|\tilde{D}^\dagger(\beta_p)\otimes I_{C}\otimes I_{R}\tilde{U}\rho_i\}|}\\
=&\sqrt{\bigg|\sum_{l=0}^{\infty}\frac{(-1)^l}{l!}\langle\beta|\tilde{U}^{\dagger}\tilde{D}(\beta_p)(\tilde{a}_{p0}^\dagger)^l(\tilde{a}_{p0})^l\tilde{D}^\dagger (\beta_p)\tilde{U}|\beta\rangle\bigg|}\\
=&\sqrt{\bigg|\sum_{l=0}^{\infty}\frac{(-1)^l}{l!}\langle\beta|e^{\beta_p\tilde{a}_{pL}^\dagger-\beta_p^*\tilde{a}_{pL}}(\tilde{a}_{pL}^\dagger)^l(\tilde{a}_{pL})^le^{-\beta_p\tilde{a}_{pL}^\dagger+\beta_p^*\tilde{a}_{pL}}|\beta\rangle\bigg|},
\end{split}
\end{align}
where $\tilde{D}(\beta_p)={\rm{exp}}(\beta_p\tilde{a}_{p0}^{\dagger}-\beta_p^*\tilde{a}_{p0})$ is the displacement operator. According to Eqs. (30) and (37), the transmittance (black dashed lines) and fidelity (red solid lines) of the single-photon coherent state under various detunings $\omega$ and dephasing rates $\gamma_0$ are plotted in Fig. 2(a) and Fig. 2(b), respectively. The fidelity of the coherent probe photons deteriorates rapidly as $\omega $ increases. Although the dephasing rate has the greater potential to negatively affect the transmittance of the probe photons, the change in fidelity as $\gamma_0$ increases is small. This difference can be explained as follows: $\gamma_0$ only attenuates the amplitude of the probe photons, whereas $\omega$ not only attenuates the amplitude but also changes its phase. Therefore, $\omega$ has a greater influence on the wave function of the photons, rendering fidelity more difficult to maintain. To enable correspondence with more commonly used parameters, we use the replacement of $\frac{g^2N}{c}=\frac{\alpha\Gamma}{4L}$, where $\alpha$ denotes the optical depth of the EIT medium~\cite{ChenQFC1, ChenQFC2}.

\FigTwo

For the input probe field in the Fock state, the density matrix of the input probe photons is expressed as $\rho^P(0,\omega)=|n_{p0}\rangle\langle n_{p0}|$. On the basis of Eq. (35), the density matrix element of the output probe photons is given by
\begin{align}
\begin{split}
\rho_{mn}^{P}&(L, \omega)\\
&={\rm{Tr}}\{\tilde{\rho}_{mn}(L,\omega)\rho_i\}\\
&=\sum_{l=0}^{\infty}\chi_{mnl}{\rm{Tr}}_P\{(c_1^{*}\tilde{a}_{p0}^{\dagger})^{n+l}(c_1\tilde{a}_{p0})^{m+l}|n_{p0}\rangle\langle n_{p0}|\},
\end{split}
\end{align}
where the disappearance of the operator $\tilde{c}_2$ is because the probe photons in the Fock state yield $\langle\tilde{a}_{p0}\rangle=0$, which makes the coefficients $\kappa_b$ and $\kappa_c$ directly related to the operator $\tilde{c}_2$ both zero, as shown in Eqs. (19) and (20). If we use the single-photon Fock state of $n_{p0}=1$ to describe the input probe photons, only two terms in Eq. (38) are nonzero: $\rho^P_{00}=1-|c_1|^2$ and $\rho^P_{11}=|c_1|^2$. Thus, the quantum state of the output probe photons becomes a mixed state, and its density matrix is $\rho^P (L,\omega)=(1-|c_1|^2)|0\rangle\langle 0|+|c_1|^2|1\rangle\langle 1|$. This shows that although the Fock-state probe field is not affected by the coupling field fluctuations, the ground-state dephasing or two-photon detuning in the EIT medium will still destroy the quantum state of the probe field and transition it into a mixed state. Similarly, the general formula for the Fock state of the input probe photons with a photon number $n_{p0}$ can be obtained by
\begin{align}
\begin{split}
\rho_{nn}^{P}&(L, \omega)\\
&=\sum_{k=n}^{n_{p0}}\chi_{nn(k-n)}\langle n_{p0}|(c_1^{*}\tilde{a}_{p0}^{\dagger})^{k}(c_1\tilde{a}_{p0})^{k}|n_{p0}\rangle\\
&=\sum_{k=n}^{n_{p0}}\chi_{nn(k-n)}\frac{n_{p0}!}{(n_{p0}-k)!}|c_1|^{2k},
\end{split}
\end{align}
by which the state of the output probe photons becomes
\begin{align}
\rho^P(L,\omega)=\sum^{n_{p0}}_{n=0}\bigg[\sum^{n_{p0}}_{k=n}\chi_{nn(k-n)}\frac{n_{p0}!}{(n_{p0}-k)!}|c_1|^{2k}\bigg]|n\rangle\langle n|.
\end{align}
Therefore, the EIT fidelity of the Fock-state probe photons $|n_{p0}\rangle$ is given by
\begin{align}
\begin{split}
F&=\sqrt{\langle n_{p0}|(\rho^P_{{n_{p0}}{n_{p0}}}|n_{p0}\rangle\langle n_{p0}|)|n_{p0}\rangle}\\
&=\sqrt{\sum_{l=0}^{\infty}\chi_{{n_{p0}}{n_{p0}}l}\langle n_{p0}|(c_1^*\tilde{a}_{p0}^\dagger)^{n_{p0}+l}(c_1\tilde{a}_{p0})^{n_{p0}+l}|n_{p0}\rangle}\\
&=\sqrt{\chi_{{n_{p0}}{n_{p0}}0}\langle n_{p0}|(c_1^*\tilde{a}_{p0}^\dagger)^{n_{p0}}(c_1\tilde{a}_{p0})^{n_{p0}}|n_{p0}\rangle}\\
&= \sqrt{\frac{1}{n_{p0}!}(|c_1|^2)^{n_{p0}}(n_{p0}!)}\\
&=|c_1|^{n_{p0}}.
\end{split}
\end{align}
The transmittance (black dashed lines) and fidelity (red solid lines) of Fock-state probe photons with $n_{p0}=1$ under various detunings $\omega$ and dephasing rates $\gamma_0 $ are plotted in Fig. 3(a) and Fig. 3(b), respectively.  The fidelity of the single-photon Fock state is attenuated to a considerably smaller extent than that of the single-photon coherent state in Fig. 2(a). This demonstrates that the phase shift caused by the detuning of the EIT medium only slightly distorts the wave function of the Fock-state probe photons. Because the phase variance of the Fock state is larger than that of the coherent state, the phase shift caused by the detuning is relatively sustainable. However, when the dephasing rate is increased, because the amplitude variance of the Fock-state probe photons is zero and smaller than that of the coherent-state probe photons, the fidelity of the single-photon Fock state decreases more rapidly.

\FigThree

\FigFour

\FigFive

\FigSix


\section{Coupling Field Fluctuations}
\label{sec:CF}

We further analyze the influence of coupling field fluctuations on the transmittance and fidelity of the probe photons under various coupling field strengths and optical depths. As mentioned in Section~\ref{sec:Fidelity}, the coupling field fluctuations do not influence the Fock-state probe photons. Therefore, here we discuss only the coherent-state probe photons $\rho^P(0,\omega)=|\beta_p\rangle\langle\beta_p|$. To quantify the effects of the coupling field fluctuations on the probe photons, we use the transmittance difference $\Delta T$ and fidelity difference $\Delta F$, which are defined as follows:
\begin{align}
&\Delta T\equiv T-\lim_{\begin{subarray}{1}
                                                \kappa_b \to 0\\
                                                \kappa_c \to 0
\end{subarray}}T,\\
&\Delta F \equiv F-\lim_{\begin{subarray}{1}
                                                \kappa_b \to 0\\
                                                \kappa_c \to 0
\end{subarray}}F.
\end{align}
Under the condition of an extremely strong coupling field, both $\kappa_b$ and $\kappa_c$ are close to zero, which means that the influence of the coupling field fluctuations ($\tilde{c}_2$) on the probe photons can be ignored. In this context, the predictions of the full quantum model and semiclassical model are consistent. Therefore, the transmittance difference $\Delta T$ and fidelity difference $\Delta F$ also underscore the differences between the full quantum and semiclassical models.

The $\Delta T$ and $\Delta F$ of the coherent-state single photons propagating in the EIT medium are plotted as a function of the dephasing rate under the two coupling field strengths in Fig. 4. Both the transmittance difference and fidelity difference start at 0, indicating that when the dephasing rate is 0, the EIT medium causes no dissipation of the probe photons. This perfect EIT condition not only completely suppresses dissipation but also prevents the influence of quantum fluctuations, including those related to the coupling field and vacuum reservoir. Because of the dissipation of most of the probe photons by the EIT medium, when the dephasing rate is sufficiently high, the differences in transmittance and fidelity decrease, and both values are close to 0. Furthermore, the peak value of the transmittance difference (i.e., the maximal contribution condition of the coupling field fluctuations) is inversely proportional to the coupling field intensity [Fig. 4(a) and Fig. 4(b)]. The stronger the coupling field, the smaller the effect of its quantum fluctuations is on the probe photons. This is in line with the semiclassical premise that strong coupling coherent light can be approximated as classical light~\cite{FleischhauerQM, DantanQM, XiaoDSP, PengQM}.

We next examine the influence of coupling field fluctuations on the probe photons at two optical depths. Although the optical depth increases from 200 in Fig. 5(a) to 1000 in Fig. 5(b), the change in the maximal influence of the coupling field fluctuations is negligible. This demonstrates that a greater optical depth is associated with greater interaction between the probe photons and the EIT medium but also causes greater dissipation of the probe photons, thereby suppressing increases in the transmittance and fidelity differences attributable to coupling field fluctuations.

The fidelity difference under various probe photon number conditions is presented in Fig. 6. Because the transmittance is the ratio of the number of input photons to the number of output photons, the transmittance difference does not change with the number of input probe photons. However, as the mean photon number of the input probe field increases from 1 in Fig. 6(a) to 10 in Fig. 6(b), the maximal fidelity difference increases substantially. According to the results in Figs. 4 and 6, a probe field with a larger mean photon number and a coupling field with a smaller intensity tend to produce larger quantum fluctuation effects. These results suggest that if the current $\Lambda$-type system is close to the coherent population trapping condition~\cite{CrayCPT}, the probe field is more strongly affected by the fluctuations of the coupling field~\cite{ChuangSqueezing}. Notably, the contributions of coherent and squeezed coupling field fluctuations are discussed in Appendixes A and B, respectively. The theoretical results show that the squeezing of the coherent coupling field enhances the influence of its fluctuations on the coherent probe photons, including transmittance and fidelity.


\section{Conclusion}

Using a full quantum approach based on mean-field expansion and the general reservoir theory, we can calculate, in a three-level $\Lambda$ medium, the quantum state of the probe photons that reveal the EIT phenomenon predicted by semiclassical theory while reflecting the influence of the fluctuations of the quantized coupling field. By studying the evolution of the quantum state of the probe photons propagating in the EIT medium, we demonstrate that under the effect of coupling field fluctuations, coherent-state probe photons transition into a mixed state wherein the amplitude and phase differ. However, the probe photons receive no contribution from the coupling field fluctuations if the expectation value of the electric field operator is 0, as is the case in the Fock state. Although the Fock-state probe photons are not affected by the coupling field fluctuations, they are destroyed by the vacuum reservoir under imperfect EIT conditions and enter a mixed state. Furthermore, our study shows that the squeezed coupling field can increase the influence of its fluctuations on the quantum state of the coherent probe photons. This implies that the EIT effect can be manipulated by controlling the quantum state properties of the coupling field. In summary, the full quantum theory in this paper can not only calculate the quantum fidelity of the EIT medium, but also has the potential to be used to investigate various quantum systems related to the EIT mechanism, such as quantum frequency converters~\cite{ChenQFC1}. This allows us to examine various quantum effects in EIT-based systems from a full quantum perspective.


\appendix

\section{Second-Order Coupling Field Fluctuations}

To calculate the influence of the coupling field fluctuations on the transmittance and fidelity of the probe photons propagating in the EIT medium, the expectation values of $\delta\tilde{a}_c$ and $\delta\tilde{a}_c^\dagger$ must be determined. According to Eq. (11), the fluctuation operator of the coupling field under the mean-field expansion can be rewritten as
\begin{align}
\delta\tilde{a}_c=\tilde{a}_c-\langle \tilde{a}_c \rangle.
\end{align}
Because the expectation value of the annihilation operator $\tilde{a}_c$ depends on the quantum state of the coupling field, the fluctuation operator $\delta\tilde{a}_c$ is also related to the quantum properties of the coupling field. For consistency with the discussion presented in the main text, we employ a coherent coupling field $|\beta_c\rangle$ to calculate the second-order fluctuations of the coupling field. The first-order fluctuations are first calculated as follows:
\begin{align}
\langle\delta\tilde{a}_c\rangle&=\langle\beta_c|(\tilde{a}_c-\langle \tilde{a}_c \rangle)|\beta_c\rangle=0,\\
\langle\delta\tilde{a}_c^\dagger\rangle&=\langle\beta_c|(\tilde{a}_c-\langle \tilde{a}_c\rangle)^\dagger|\beta_c\rangle=0.
\end{align}
For the coherent coupling field, the first-order fluctuations are all zero. Next, we calculate the second-order fluctuations:
\begin{align}
\begin{split}
\langle\delta\tilde{a}_c^\dagger\delta\tilde{a}_c\rangle&=\langle\beta_c| (\tilde{a}_c^\dagger-\langle\tilde{a}_c\rangle^*)(\tilde{a}_c-\langle\tilde{a}_c\rangle)|\beta_c\rangle\\
&=\langle\tilde{a}_c^\dagger\tilde{a}_c\rangle-\langle\tilde{a}_c\rangle^*\langle\tilde{a}_c\rangle-\langle\tilde{a}_c\rangle\langle\tilde{a}_c^\dagger\rangle+|\langle\tilde{a}_c\rangle|^2\\
&=|\beta_c|^2-|\beta_c|^2-|\beta_c|^2+|\beta_c|^2\\
&=0,
\end{split}
\end{align}
\begin{align}
\begin{split}
\langle\delta\tilde{a}_c\delta\tilde{a}_c\rangle&=\langle\beta_c|(\tilde{a}_c-\langle\tilde{a}_c\rangle)(\tilde{a}_c-\langle\tilde{a}_c\rangle)|\beta_c\rangle\\
&=\langle\tilde{a}_c\tilde{a}_c\rangle-\langle\tilde{a}_c\rangle\langle\tilde{a}_c\rangle-\langle\tilde{a}_c\rangle\langle\tilde{a}_c\rangle+\langle\tilde{a}_c\rangle^2\\
&=\beta_c^2-\beta_c^2-\beta_c^2+\beta_c^2\\
&=0,
\end{split}
\end{align}
\begin{align}
\begin{split}
\langle\delta\tilde{a}_c^\dagger\delta\tilde{a}_c^\dagger\rangle&=\langle\beta_c|(\tilde{a}_c^\dagger-\langle\tilde{a}_c\rangle^*)(\tilde{a}_c^\dagger-\langle\tilde{a}_c\rangle^*)|\beta_c\rangle\\
&=\langle\tilde{a}_c^\dagger\tilde{a}_c^\dagger\rangle-\langle\tilde{a}_c\rangle^*\langle\tilde{a}_c^\dagger\rangle-\langle\tilde{a}_c\rangle^*\langle\tilde{a}_c^\dagger\rangle+(\langle\tilde{a}_c\rangle^*)^2\\
&=(\beta_c^*)^2-(\beta_c^*)^2-(\beta_c^*)^2+(\beta_c^*)^2\\
&=0,
\end{split}
\end{align}
\begin{align}
\begin{split}
\langle\delta\tilde{a}_c\delta\tilde{a}_c^\dagger\rangle&=\langle\beta_c|(\tilde{a}_c-\langle\tilde{a}_c\rangle)(\tilde{a}_c^\dagger-\langle\tilde{a}_c\rangle^*)|\beta_c\rangle\\
&=\langle\tilde{a}_c\tilde{a}_c^\dagger\rangle-\langle\tilde{a}_c\rangle^*\langle\tilde{a}_c\rangle-\langle\tilde{a}_c\rangle\langle\tilde{a}_c^\dagger\rangle+|\langle\tilde{a}_c\rangle|^2\\
&=\langle\beta_c|([\tilde{a}_c^\dagger,\tilde{a}_c]+\tilde{a}_c^\dagger\tilde{a}_c)|\beta_c\rangle-\langle\tilde{a}_c\rangle^*\langle\tilde{a}_c\rangle-\langle\tilde{a}_c\rangle\langle\tilde{a}_c^\dagger\rangle+|\langle\tilde{a}_c\rangle|^2\\
&=1+|\beta_c|^2-|\beta_c|^2-|\beta_c|^2+|\beta_c|^2\\
&=1.
\end{split}
\end{align}
Among these four second-order fluctuations, the only nonzero term, $\langle\delta\tilde{a}_c\delta\tilde{a}_c^\dagger\rangle$, affects the transmittance and fidelity of the probe field. Moreover, under the EIT condition of the strong coupling field discussed herein, the coefficients $\kappa_b$ and $\kappa_c$ accompanying the coupling field fluctuation operator $\tilde{c_2}$ in Eq. (26) are considerably smaller than 1, making the high-order (above the third order) fluctuations extremely small and therefore negligible.


\section{Squeezed Coupling Field}

Consider a scenario where the coupling field is in a squeezed coherent state, expressed as $|\xi_c, \beta_c\rangle=\tilde{D}(\beta_c)\tilde{S}(\xi_c)|0\rangle$, where $\tilde{D}(\beta_c)={\rm{exp}}(\beta_c\tilde{a}_c^\dagger-\beta_c^*\tilde{a}_c)$ is the displacement operator, $\tilde{S}(\xi_c)={\rm{exp}}\left[\frac{1}{2}(\xi_c^*\tilde{a}_c^2-\xi_c\tilde{a}_c^{\dagger 2})\right]$ is the squeeze operator, and $|0\rangle$ is the vacuum state. The first-order fluctuation is expressed as follows:
\begin{align}
\begin{split}
\langle\delta\tilde{a}_c\rangle&=\langle\xi_c, \beta_c|(\tilde{a}_c-\langle\tilde{a}_c\rangle)|\xi_c, \beta_c\rangle=0.
\end{split}
\end{align}
The same approach can be applied to the proof of $\langle\delta\tilde{a}_c^\dagger\rangle=0$. The first-order fluctuations of the squeezed coupling field are all 0,  as is the case with the coherent coupling field. However, the second-order fluctuations of the squeezed state and coherent state differ:
\begin{align}
\begin{split}
\langle\delta\tilde{a}_c^\dagger\delta\tilde{a}_c\rangle&=\langle\xi_c, \beta_c|(\tilde{a}_c^\dagger-\langle\tilde{a}_c\rangle^*)(\tilde{a}_c-\langle\tilde{a}_c\rangle)|\xi_c, \beta_c\rangle\\
&=\langle\tilde{a}_c^\dagger\tilde{a}_c\rangle-\langle\tilde{a}_c\rangle^*\langle\tilde{a}_c\rangle-\langle\tilde{a}_c\rangle\langle\tilde{a}_c^\dagger\rangle+|\langle\tilde{a}_c\rangle|^2\\
&=|\beta_c|^2+{\rm{sinh}}^2r_c-|\beta_c|^2-|\beta_c|^2+|\beta_c|^2\\
&={\rm{sinh}}^2r_c,
\end{split}
\end{align}
\begin{align}
\begin{split}
\langle\delta\tilde{a}_c\delta\tilde{a}_c\rangle&=\langle\xi_c, \beta_c|(\tilde{a}_c-\langle\tilde{a}_c\rangle)(\tilde{a}_c-\langle\tilde{a}_c\rangle)|\xi_c, \beta_c\rangle\\
&=\langle\tilde{a}_c\tilde{a}_c\rangle-\langle\tilde{a}_c\rangle\langle\tilde{a}_c\rangle-\langle\tilde{a}_c\rangle\langle\tilde{a}_c\rangle+\langle\tilde{a}_c\rangle^2\\
&=\beta_c^2+\frac{1}{2}{\rm{sinh}}(2r_c)e^{i\theta_c}-\beta_c^2-\beta_c^2+\beta_c^2\\
&=\frac{1}{2}{\rm{sinh}}(2r_c)e^{i\theta_c},
\end{split}
\end{align}
\begin{align}
\begin{split}
\langle\delta\tilde{a}_c^\dagger\delta\tilde{a}_c^\dagger\rangle&=\langle\xi_c, \beta_c|(\tilde{a}_c^\dagger-\langle\tilde{a}_c\rangle^*)(\tilde{a}_c^\dagger-\langle\tilde{a}_c\rangle^*)|\xi_c, \beta_c\rangle\\
&=\langle\tilde{a}_c^\dagger\tilde{a}_c^\dagger\rangle-\langle\tilde{a}_c\rangle^*\langle\tilde{a}_c^\dagger\rangle-\langle\tilde{a}_c\rangle^*\langle\tilde{a}_c^\dagger\rangle+(\langle\tilde{a}_c\rangle^*)^2\\
&=(\beta_c^*)^2+\frac{1}{2}{\rm{sinh}}(2r_c)e^{-i\theta_c}-(\beta_c^*)^2-(\beta_c^*)^2+(\beta_c^*)^2\\
&=\frac{1}{2}{\rm{sinh}}(2r_c)e^{-i\theta_c},
\end{split}
\end{align}
\begin{align}
\begin{split}
\langle\delta\tilde{a}_c\delta\tilde{a}_c^\dagger\rangle&=\langle\xi_c, \beta_c|(\tilde{a}_c-\langle\tilde{a}_c\rangle)(\tilde{a}_c^\dagger-\langle\tilde{a}_c\rangle^*)|\xi_c, \beta_c\rangle\\
&=\langle\tilde{a}_c\tilde{a}_c^\dagger\rangle-\langle\tilde{a}_c\rangle^*\langle\tilde{a}_c\rangle-\langle\tilde{a}_c\rangle\langle\tilde{a}_c^\dagger\rangle+|\langle\tilde{a}_c\rangle|^2\\
&=\langle\xi_c, \beta_c|([\tilde{a}_c^\dagger,\tilde{a}_c]+\tilde{a}_c^\dagger\tilde{a}_c)|\xi_c, \beta_c\rangle-\langle\tilde{a}_c\rangle^*\langle\tilde{a}_c\rangle-\langle\tilde{a}_c\rangle\langle\tilde{a}_c^\dagger\rangle+|\langle\tilde{a}_c\rangle|^2\\
&=1+|\beta_c|^2+{\rm{sinh}}^2r_c-|\beta_c|^2-|\beta_c|^2+|\beta_c|^2\\
&=1+{\rm{sinh}}^2r_c,
\end{split}
\end{align}
where $\xi_c=r_{c}e^{i\theta_c}$, and where $r$ is the squeeze parameter. Because the hyperbolic sine function increases exponentially with the squeeze parameter, the squeezing increases the second-order fluctuations, which become dominant as the degree of squeezing increases. This indicates that the influence of quantum fluctuations on the probe photons propagated in the EIT medium can be enhanced by a squeezed coupling field.


\section*{ACKNOWLEDGMENTS}

We thank You-Lin Chuang for helpful discussions. This work was supported by Taiwan's Ministry of Science and Technology (Grant No. 110-2112-M-006-014).


\end{document}